\pdfoutput=1
\documentclass[a4paper,11pt]{article}
\usepackage{pos}
\usepackage{multirow}

\title{Multi-messenger searches via IceCube’s high-energy neutrinos and gravitational-wave detections of LIGO/Virgo}
 \ShortTitle{IceCube GWTC-2 analysis}

\author{The IceCube Collaboration \\{\normalsize \normalfont(a complete list of authors can be found at the end of the proceedings)}}

\emailAdd{dv2397@columbia.edu}

\abstract{We summarize initial results for high-energy neutrino counterpart searches coinciding with gravitational-wave events in LIGO/Virgo's GWTC-2 catalog using IceCube's neutrino triggers. We did not find any statistically significant high-energy neutrino counterpart and derived upper limits on the time-integrated neutrino emission on Earth as well as the isotropic equivalent energy emitted in high-energy neutrinos for each event.

\vspace{4mm}
{\bfseries Corresponding authors:}
Do\u{g}a Veske$^{1*}$, 
Raamis Hussain$^{2}$, 
Zsuzsa M\'arka$^{1}$, 
Stefan Countryman$^{1}$, 
Alex Pizzuto$^{2}$, 
Yasmeen Asali$^{1,3}$, 
Ana Silva Oliveira$^{1}$ and
Justin Vandenbroucke$^{2}$\\
{$^{1}$ \itshape Columbia University, NY, USA}\\
{$^{2}$ \itshape University of Wisconsin Madison, WI, USA}\\
{$^{3}$ \itshape Yale University, CT, USA}\\[4mm]
$^*$ Presenter

\FullConference{37$^{\rm{th}}$ International Cosmic Ray Conference (ICRC 2021)\\
		July 12th -- 23rd, 2021\\
		Online -- Berlin, Germany}

}

\begin{document}
\maketitle

\section{Introduction}\label{sec:info}

Astrophysics presents us the opportunity to observe physical phenomena which are not feasibly created via experiments on Earth. Observing such astrophysical events with multiple messengers can extend our understanding by showing a more complete picture of the emission processes from the astrophysical source. With developing detectors, platforms for communication between astronomy communities, and efficient statistical methods \citep{Veske_2021}, multi-messenger searches and detections \citep{2017ApJ...848L..12A,Monitor:2017mdv,ic1709022mm} have become a reality. One multi-messenger combination that hasn't been observed yet is of gravitational-waves (GWs) and high-energy neutrinos, despite previous searches \citep{ 2008CQGra..25k4039A, 2009IJMPD..18.1655V, 2011PhRvL.107y1101B, 2012PhRvD..85j3004B, 2013JCAP...06..008A, 2014PhRvD..90j2002A, 2016PhRvD..93l2010A, 2017ApJ...850L..35A,2017PhRvD..96b2005A, Aartsen:2020mla}. Here we present our searches for high-energy neutrinos originating from the sources of GW events in the GWTC-2 catalog of the LIGO Scientific and Virgo Collaborations \citep{Abbott:2020niy} with the triggers of the IceCube detector \citep{Aartsen:2016nxy}. The methodology of the searches are summarized in Sec. \ref{sec:searches} and the results are given in Sec. \ref{sec:results}. Sec. \ref{sec:conc} has the conclusion.

\subsection{First half of the third observing run of advanced LIGO and Virgo}
First half of the third observing run of advanced LIGO \citep{TheLIGOScientific:2014jea} and Virgo \citep{,TheVirgo:2014hva} detectors, which is generally referred as O3a, started on April 1$^{\rm st}$, 2019 and ended six months later on October 1$^{\rm st}$. It was a combined run of the two LIGO detectors in the US (Hanford and Livingston) and the Virgo detector in Italy. During the run, candidate events were announced through Gravitational-wave Candidate Event Database (GraceDB)\footnote{\url{https://gracedb.ligo.org}} in low-latency. The IceCube Collaboration performed realtime follow-up of these open public alerts \citep{countryman2019lowlatency,Keivani:2019Mx,Hussain:2019gQ} and sent out Gamma-ray Coordinates Network circulars for our findings. After the offline analysis by LIGO Scientific and Virgo Collaborations, the second gravitational-wave transients catalog GWTC-2 was released with total of 39 events of coalescing binary compact objects from O3a \citep{Abbott:2020niy}. Twenty-six of these events were previously reported in low-latency during O3a, while thirteen events were new detections found in the offline analysis of the O3a data.

\section{Searches}
\label{sec:searches}
There are two pipelines used for the searches; Low Latency Algorithm for Multi-messenger Astrophysics (LLAMA) and the Unbinned Maximum Likelihood (UML). Although both pipelines are essentially looking for the same thing, there are differences between them. Each search uses a different statistical approach to search for and quantify the significance of IceCube neutrinos coincident with compact binary mergers. We briefly describe both methods below. For details see Ref. \cite{PhysRevD.100.083017} for the LLAMA search method and Ref. \cite{Hussain:2019gQ} for the UML method. Both pipelines use the gamma-ray follow up (GFU) stream provided by IceCube \citep{Kintscher:2016uqh} for high-energy neutrino data. This data sample mostly consists of atmospheric muons with a small portion of astrophysical neutrino triggers.

\subsection{Low Latency Algorithm for Multi-messenger Astrophysics}

The LLAMA search calculates the Bayesian probabilities of different hypotheses arising from the combinations of GW or neutrinos being astrophysically related or not, considering the scenario of them being not astrophysical at all as well. The odds ratio of the GW and neutrino messengers being astrophysically related compared to all the other combinations is used as the test statistic. The priors on the probabilities are either obtained from the detectors' background trigger rate due to noise or by assuming an astrophysical energy emission distribution for GW and neutrinos. In the case for certain GW detections, as in here, likelihoods for messengers' origin (noise or astrophysical) and their relation are calculated by using their detection times, sky localizations, reconstructed energy of neutrinos and the estimated luminosity distance of the GW event. Considering the pipelines here, using the distance of the GW event is unique to the LLAMA search which is used to account for the propagation of the neutrinos in space. The maximum allowed time difference between related GW and neutrinos is $\pm500$~s \citep{BARET20111} and temporally closer detections are favored.

The significances are obtained by using precomputed background distributions which contain the results of randomly matched neutrinos and simulated GWs at the detection rate of the GFU stream. For each type of merger (binary black hole (BBH), neutron star black hole (NSBH) or binary neutron star (BNS)) different background distributions are obtained. The reason for this is the different detection horizon of the GW detectors for each source type due to the different signal power created by them in the detectors. This produces different distance distributions for each type of merger which affects the significance since the distance information is used in the test statistic calculation.

\subsection{Unbinned Maximum Likelihood}
This method uses an unbinned maximum likelihood (UML) which is weighted by a spatial weight derived from the sky localization of the GWs. The method is briefly described here, but for full details on the method, see \cite{Aartsen:2020mla}.

Firstly, the sky is divided into equal area bins using the Healpix pixelization scheme. The test statistic is then calculated in every pixel by maximizing the log-likelihood ratio with respect to the number of signal events, $n_s$, and the spectral index of the source, $\gamma$. The test statistic in each pixel is then weighted by the spatial weight which describes the probability of the GW source being located in the given pixel. The maximum test statistic in the sky is chosen as the best fit location for the scan.

To compute the significance of a given observation, we run 30,000 trials for each GW event with scrambled neutrino data to build a background test statistic distribution. We then compare the observed test statistic to the background distribution to compute a p-value for each GW.

Two analyses are performed using the UML method. The first is a short timescale follow up of every reported GW event. Here we search for neutrinos within a $\pm$500~s time window centered around the GW merger time. The goal of this analysis is to search for prompt neutrino emission just before and just after the GW merger. This analysis was run in real-time during the O3 observing run and responded to all public alerts sent by LVC in low-latency.

The second analysis is a longer time scale search targeting all binary neutron star and neutron star-black hole mergers. Here we search a [-0.1,+14]~day time window around the GW merger time. This search is motivated by several models which predict neutrino emission specifically from BNS and NSBH mergers on longer time scales \cite{Fang:2017tla,Decoene:2019eux}. This analysis was run on three events from GWTC-2: GW190425, GW190426\_152155, and GW190814 which contain at least one compact object less massive than 3 M$_\odot$ which could be a neutron star.

\section{Results}
\label{sec:results}
\begin{figure}[h]
    \centering
    \includegraphics[width=0.7\textwidth]{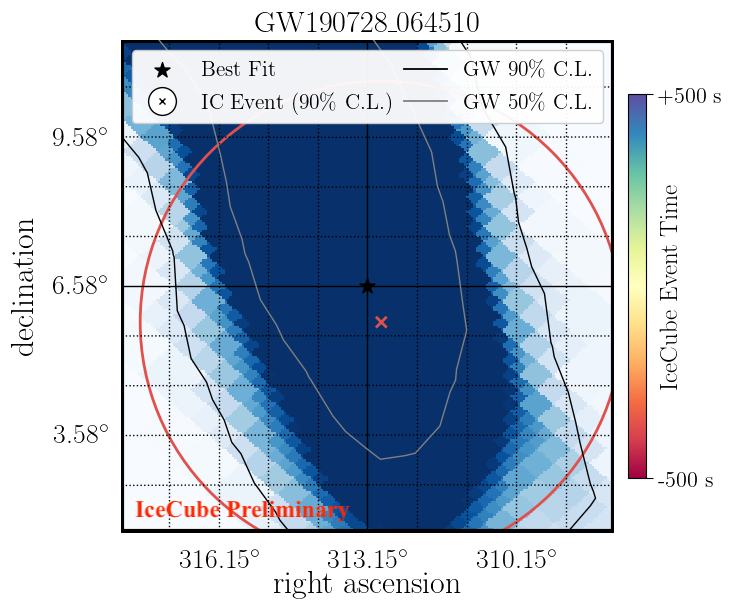}
    \caption{The zoomed in display of the sky localization of the most significant high-energy neutrino-GW pair in equatorial coordinates. The blue gradient represents the sky localization probability of the GW with the darker color representing higher probability. The red cross represents the best fit direction for the coincident neutrino with the circle representing the 90\% containment angular error region. The figure is centered around the best fit location of the UML search, which is shown with a black star at the center.}
    \label{fig:gw190728}
\end{figure}

\begin{figure}[h]
    \centering
    \includegraphics[width=0.8\textwidth]{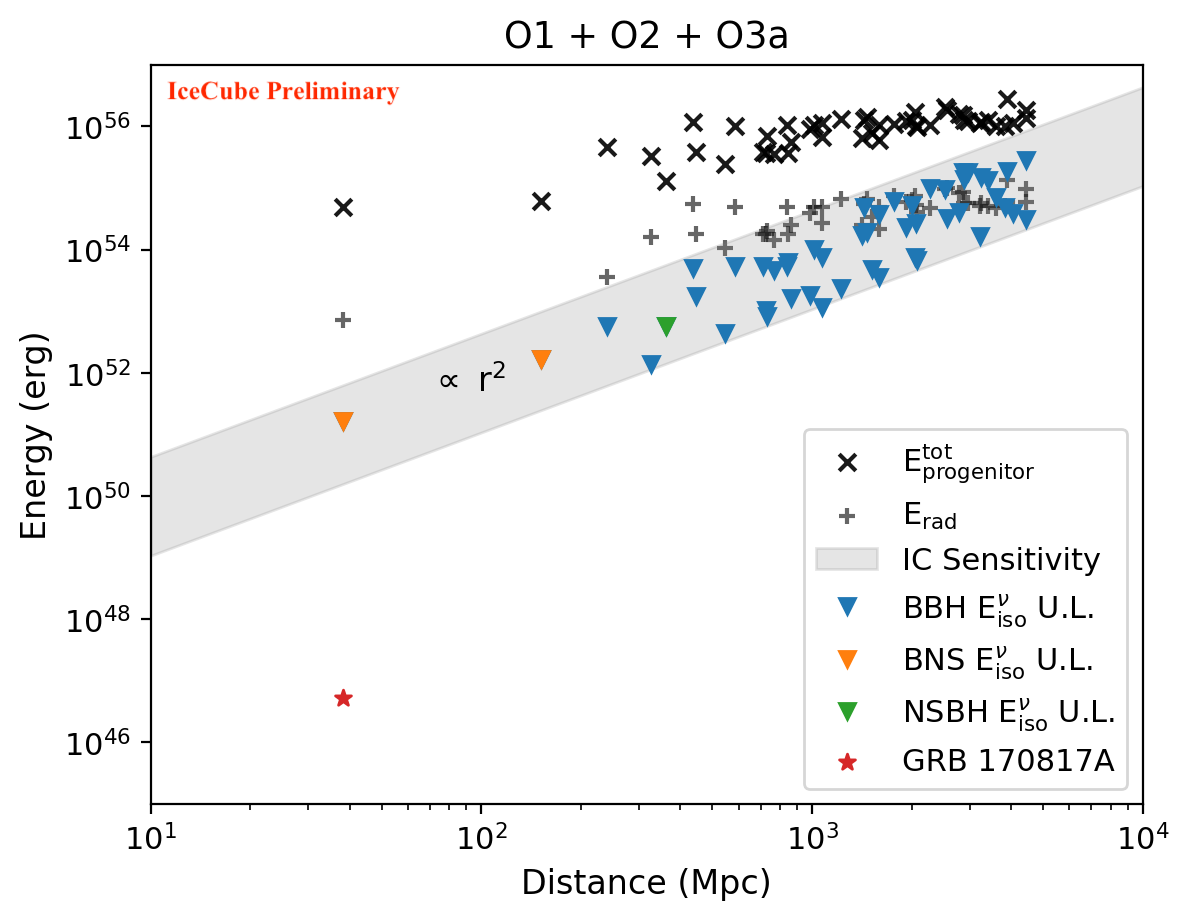}
    \caption{90\% UL on the  isotropic equivalent energy emitted in high-energy neutrinos during a 1000~s time window (blue and orange triangles). $E_{\mathrm{progenitor}}^{\mathrm{tot}}$ (black cross) is the total rest mass energy of the progenitors and $E_{\mathrm{rad}}$ (gray cross) is the total radiated energy of the binary system. While not all of the progenitor energy is available for acceleration processes, we show it here as a relevant energy scale in the binary system. The median distances for each GW are taken from GWTC-1 and GWTC-2 \citep{LIGOScientific:2018mvr,Abbott:2020niy}. Note that errors on the distance measurements are significant but not shown here for clarity. The red star shows the measured $E_{\mathrm{iso}}$ for GRB 170817A by $Fermi$ GBM taken from \cite{Monitor:2017mdv}. The gray band represents the range of 90\% $E_{\mathrm{iso}}$ ULs that IceCube can set based on the range of point source sensitivities.}
    \label{fig:eiso}
\end{figure}

Among the events in GWTC-2 no significant neutrino emission was observed in either the UML or LLAMA analyses. The long time scale searches with the UML also yielded no significant results. Table \ref{tab:results} shows the full results for the 1000~s follow up of the events from the GWTC-2 catalog. The table contains the 90\% confidence level upper limits (UL) computed for the energy scaled time-integrated neutrino flux (E${^2}$F) and the isotropic equivalent energy (E$_{\mathrm{iso}}$) emitted in high-energy neutrinos from each GW event.
Table \ref{tab:2week_GWTC2} shows the results for the 2~week follow up of BNS/NSBH candidates from GWTC-2.
Fig. \ref{fig:eiso} shows the E$_{\mathrm{iso}}$ upper limits as a function of the distance to the source including the events from GWTC-1 \citep{LIGOScientific:2018mvr} as well. The overlay of the neutrino and the zoomed in GW sky localization for the most significant event GW190728\_064510 which has a $p$-value of 1.3\% and 4\% in the LLAMA and UML pipelines respectively is shown in Fig. \ref{fig:gw190728}. The candidate event of GW190728\_064510 (S190728q) was also the only event that had $\leq$1\% $p$-value in at least one of the two analyses during the realtime follow-up. The coordinates for the signficant neutrino were shared via GCN circular \footnote{\url{https://gcn.gsfc.nasa.gov/gcn3/25210.gcn3}}. The coincident neutrino arrived 360~s before the GW merger and had a reconstructed energy of 601~GeV. No additional counterpart was found from other observatories (i.e. Ref. \citep{Keivani_2021}).

\begin{table}[]
    \centering
    \resizebox{0.85\textwidth}{!}{\rowcolors{1}{}{gray!25}
    \begin{tabular}{|c|c|cc|ccc|}
    \hline
    \multicolumn{2}{|c}{  }&\multicolumn{2}{|c}{ LLAMA } &\multicolumn{3}{|c|}{UML} \\ \hline
    \rowcolor{gray!0.01} &  &  & E${^2}$F UL &  & E${^2}$F UL &  \\
    \multirow{-2}{*}{Event} & \multirow{-2}{*}{Type} & \multirow{-2}{*}{$p$-value} & [GeVcm$^{-2}$] & \multirow{-2}{*}{$p$-value} & [GeVcm$^{-2}$] & \multirow{-2}{*}{E$_{\rm iso}$ UL [erg]}\\ \hline
    GW190408\_181802  & BBH  & 0.16  & 0.048 & 0.17 & 0.0512 &  4.85 $\times$ 10$^{53}$ \\ \hline
    GW190412          & BBH  & 0.19  & 0.041 & 0.13 & 0.0459 &  8.31 $\times$ 10$^{52}$ \\ \hline
    GW190413\_052954  & BBH  & 0.21  & 0.087 & 0.28 & 0.133  &  7.01 $\times$ 10$^{54}$ \\ \hline
    GW190413\_134308  & BBH  & 0.18  & 0.34  & 0.34 & 0.270  &  2.84 $\times$ 10$^{55}$ \\ \hline
    GW190421\_213856  & BBH  & 0.77  & 0.46  & 0.56 & 0.393  &  1.40 $\times$ 10$^{55}$ \\ \hline
    GW190424\_180648  & BBH  & 0.58  & 0.32  & 0.23 & 0.233  &  5.37 $\times$ 10$^{54}$ \\ \hline
    GW190425          & BNS  & 0.16  & 0.22  & 0.94 & 0.176  &  1.66 $\times$ 10$^{52}$ \\ \hline
    GW190426\_152155  & NSBH & 0.12  & 0.082 & 0.12 & 0.0942 &  5.65 $\times$ 10$^{52}$ \\ \hline
    GW190503\_185404  & BBH  & 0.87  & 0.54  & 0.34 & 0.584  &  4.99 $\times$ 10$^{54}$ \\ \hline
    GW190512\_180714  & BBH  & 0.67  & 0.23  & 0.85 & 0.199  &  1.74 $\times$ 10$^{54}$ \\ \hline
    GW190513\_205428  & BBH  & 0.97  & 0.043 & 0.94 & 0.0514 &  6.73 $\times$ 10$^{53}$ \\ \hline
    GW190514\_065416  & BBH  & 0.28  & 0.089 & 0.44 & 0.0453 &  3.96 $\times$ 10$^{54}$ \\ \hline
    GW190517\_055101  & BBH  & 0.14  & 0.48  & 0.26 & 0.366  &  6.05 $\times$ 10$^{54}$ \\ \hline
    GW190519\_153544  & BBH  & 0.063 & 0.15  & 0.21 & 0.0914 &  3.20 $\times$ 10$^{54}$ \\ \hline
    GW190521          & BBH  & 0.47  & 0.37  & 0.63 & 0.359  &  1.90 $\times$ 10$^{55}$ \\ \hline
    GW190521\_074359  & BBH  & 0.16  & 0.049 & 0.15 & 0.0451 &  2.36 $\times$ 10$^{53}$ \\ \hline
    GW190527\_092055  & BBH  & 0.61  & 0.41  & 0.88 & 0.326  &  1.01 $\times$ 10$^{55}$ \\ \hline
    GW190602\_175927  & BBH  & 0.22  & 0.34  & 0.17 & 0.370  &  9.73 $\times$ 10$^{54}$ \\ \hline
    GW190620\_030421  & BBH  & 0.15  & 0.36  & 0.23 & 0.121  &  4.13 $\times$ 10$^{54}$ \\ \hline
    GW190630\_185205  & BBH  & 0.38  & 0.15  & 0.81 & 0.427  &  5.31 $\times$ 10$^{53}$ \\ \hline
    GW190701\_203306  & BBH  & 1.0   & 0.039 & 0.87 & 0.0385 &  7.65 $\times$ 10$^{53}$ \\ \hline
    GW190706\_222641  & BBH  & 0.99  & 0.036 & 0.92 & 0.0356 &  3.17 $\times$ 10$^{54}$ \\ \hline
    GW190707\_093326  & BBH  & 0.43  & 0.24  & 0.63 & 0.202  &  4.74 $\times$ 10$^{53}$ \\ \hline
    GW190708\_232457  & BBH  & 0.11  & 0.11  & 0.56 & 0.0720 &  1.62 $\times$ 10$^{53}$ \\ \hline
    GW190719\_215514  & BBH  & 0.79  & 0.054 & 0.91 & 0.0512 &  4.90 $\times$ 10$^{54}$ \\ \hline
    GW190720\_000836  & BBH  & 0.98  & 0.13  & 0.94 & 0.0872 &  5.34 $\times$ 10$^{53}$ \\ \hline
    GW190727\_060333  & BBH  & 0.79  & 0.38  & 0.74 & 0.324  &  1.53 $\times$ 10$^{55}$ \\ \hline
    GW190728\_064510  & BBH  & 0.013 & 0.89  & 0.04 & 0.315  &  6.36 $\times$ 10$^{53}$ \\ \hline
    GW190731\_140936  & BBH  & 0.29  & 0.93  & 0.61 & 0.385  &  1.81 $\times$ 10$^{55}$ \\ \hline
    GW190803\_022701  & BBH  & 0.21  & 0.037 & 0.64 & 0.0354 &  1.69 $\times$ 10$^{54}$ \\ \hline
    GW190814          & BBH  & 1.0   & 0.24  & 1.0  & 0.259  &  5.68 $\times$ 10$^{52}$ \\ \hline
    GW190828\_063405  & BBH  & 0.86  & 0.21  & 0.98 & 0.178  &  2.74 $\times$ 10$^{54}$ \\ \hline
    GW190828\_065509  & BBH  & 0.72  & 0.38  & 0.84 & 0.368  &  3.73 $\times$ 10$^{54}$ \\ \hline
    GW190909\_114149  & BBH  & 0.56  & 0.11  & 0.39 & 0.136  &  1.33 $\times$ 10$^{55}$ \\ \hline
    GW190910\_112807  & BBH  & 0.16  & 0.45  & 0.77 & 0.177  &  1.90 $\times$ 10$^{54}$ \\ \hline
    GW190915\_235702  & BBH  & 0.40  & 0.036 & 0.44 & 0.0354 &  3.61 $\times$ 10$^{53}$ \\ \hline
    GW190924\_021846  & BBH  & 0.038 & 0.037 & 0.23 & 0.0346 &  4.46 $\times$ 10$^{52}$ \\ \hline
    GW190929\_012149  & BBH  & 0.091 & 0.34  & 0.22 & 0.276  &  -- \\ \hline
    GW190930\_133541  & BBH  & 0.19  & 0.038 & 0.31 & 0.0427 &  1.05 $\times$ 10$^{53}$ \\ \hline
\end{tabular}}
\caption{Results for the events in GWTC-2 for the 1000~s follow up. GW190814 is labelled as a BBH merger here although the type of the lighter object at $\sim2.6$ M$_\odot$ is unknown  \cite{Abbott:2020khf}. Note that the $E_{\mathrm{iso}}$ UL for GW190929\_012149 could not be properly calculated due to the distance measure being undefined for many pixels in the skymap, so it is omitted here.}
\label{tab:results}
\end{table}

\begin{table}[]
    \centering
    \resizebox{0.8\textwidth}{!}{
    \begin{tabular}{|c|c|cc|ccc|}
    \hline
  Event & Type &  $p$-value &  E${^2}$F UL [GeVcm$^{-2}$] \\ \hline
  GW190425         & BNS   &   0.43 & 0.661  \\ \hline
  GW190426\_152155  & NSBH  &   0.21 & 0.248  \\ \hline
  GW190814         & BBH   &   0.59 & 0.309  \\ \hline
    \end{tabular}}
    \caption{Results for the 2 week follow up analysis using the UML method. These 3 events from GWTC-2 were followed up as they were the only potential BNS/NSBH candidates.}
    \label{tab:2week_GWTC2}
\end{table}

\newpage
\section{Conclusion}
\label{sec:conc}
We summarized our searches for high-energy neutrino counterparts to GW events and their results for the events in the GWTC-2 catalog. Searches were done by two pipelines, LLAMA and UML, and neither of them found any statistically significant event. We also derived upper limits on the neutrino emission fluence on Earth and isotropically emitted energy in high-energy neutrinos. The searches for high-energy neutrino counterparts to GW events with the neutrino triggers of IceCube will continue as new GW catalogs are released.
\section*{Acknowledgements}
The Columbia Experimental Gravity group is grateful for the generous support
of the National Science Foundation under grant PHY-2012035.

\bibliographystyle{ICRC}
\bibliography{references}

\providecommand{\href}[2]{#2}\begingroup\raggedright\begin{thebibliography}{10}

\bibitem{Veske_2021}
D.~Veske, Z.~M{\'{a}}rka, I.~Bartos, and S.~M{\'{a}}rka
  \href{http://dx.doi.org/10.3847/1538-4357/abd542}{{\em The Astrophysical
  Journal} {\bfseries 908} no.~2, (Feb, 2021) 216}.

\bibitem{2017ApJ...848L..12A}
B.~P. {Abbott} {\em et~al.}
  \href{http://dx.doi.org/10.3847/2041-8213/aa91c9}{{\em The Astrophysical
  Journal Letters} {\bfseries 848} (Oct., 2017) L12}.

\bibitem{Monitor:2017mdv}
{\bfseries LIGO Scientific, Virgo, Fermi-GBM, INTEGRAL} Collaboration, B.~P.
  Abbott {\em et~al.} \href{http://dx.doi.org/10.3847/2041-8213/aa920c}{{\em
  Astrophys. J.} {\bfseries 848} no.~2, (2017) L13}.

\bibitem{ic1709022mm}
M.~G. Aartsen {\em et~al.} {\em Science} {\bfseries 361} (July, 2018) .

\bibitem{2008CQGra..25k4039A}
Y.~{Aso} {\em et~al.}
  \href{http://dx.doi.org/10.1088/0264-9381/25/11/114039}{{\em Class. Quantum
  Grav} {\bfseries 25} no.~11, (June, 2008) 114039}.

\bibitem{2009IJMPD..18.1655V}
V.~{van Elewyck} {\em et~al.}
  \href{http://dx.doi.org/10.1142/S0218271809015655}{{\em Int. J. Mod. Phys. D}
  {\bfseries 18} (2009) 1655--1659}.

\bibitem{2011PhRvL.107y1101B}
I.~{Bartos}, C.~{Finley}, A.~{Corsi}, and S.~{M{\'a}rka}
  \href{http://dx.doi.org/10.1103/PhysRevLett.107.251101}{{\em Phys.Rev.Lett.}
  {\bfseries 107} no.~25, (Dec., 2011) 251101}.

\bibitem{2012PhRvD..85j3004B}
B.~{Baret} {\em et~al.}
  \href{http://dx.doi.org/10.1103/PhysRevD.85.103004}{{\em Phys.Rev.D}
  {\bfseries 85} no.~10, (May, 2012) 103004}.

\bibitem{2013JCAP...06..008A}
S.~{Adri{\'a}n-Mart{\'{\i}}nez} {\em et~al.}
  \href{http://dx.doi.org/10.1088/1475-7516/2013/06/008}{{\em JCAP} {\bfseries
  6} (June, 2013) 008}.

\bibitem{2014PhRvD..90j2002A}
M.~G. {Aartsen} {\em et~al.}
  \href{http://dx.doi.org/10.1103/PhysRevD.90.102002}{{\em Phys.Rev.D}
  {\bfseries 90} no.~10, (Nov., 2014) 102002}.

\bibitem{2016PhRvD..93l2010A}
S.~{Adri{\'a}n-Mart{\'{\i}}nez} {\em et~al.}
  \href{http://dx.doi.org/10.1103/PhysRevD.93.122010}{{\em Phys.Rev.D}
  {\bfseries 93} no.~12, (June, 2016) 122010}.

\bibitem{2017ApJ...850L..35A}
A.~{Albert} {\em et~al.} \href{http://dx.doi.org/10.3847/2041-8213/aa9aed}{{\em
  Astrophys.J.Lett.} {\bfseries 850} (Dec., 2017) L35}.

\bibitem{2017PhRvD..96b2005A}
A.~{Albert} {\em et~al.}
  \href{http://dx.doi.org/10.1103/PhysRevD.96.022005}{{\em Phys.Rev.D}
  {\bfseries 96} no.~2, (July, 2017) 022005}.

\bibitem{Aartsen:2020mla}
{\bfseries IceCube} Collaboration, M.~G. Aartsen {\em et~al.}
  \href{http://dx.doi.org/10.3847/2041-8213/ab9d24}{{\em Astrophys. J. Lett.}
  {\bfseries 898} no.~1, (2020) L10}.

\bibitem{Abbott:2020niy}
{\bfseries {LIGO Scientific} and {Virgo}} Collaboration, R.~Abbott {\em et~al.}
  \href{http://dx.doi.org/10.1103/PhysRevX.11.021053}{{\em Phys. Rev. X}
  {\bfseries 11} (Jun, 2021) 021053}.

\bibitem{Aartsen:2016nxy}
{\bfseries IceCube} Collaboration, M.~G. Aartsen {\em et~al.}
  \href{http://dx.doi.org/10.1088/1748-0221/12/03/P03012}{{\em JINST}
  {\bfseries 12} no.~03, (2017) P03012}.

\bibitem{TheLIGOScientific:2014jea}
{\bfseries LIGO Scientific} Collaboration, J.~Aasi {\em et~al.}
  \href{http://dx.doi.org/10.1088/0264-9381/32/7/074001}{{\em Class. Quant.
  Grav.} {\bfseries 32} (2015) 074001}.

\bibitem{TheVirgo:2014hva}
{\bfseries Virgo} Collaboration, F.~Acernese {\em et~al.}
  \href{http://dx.doi.org/10.1088/0264-9381/32/2/024001}{{\em Class. Quant.
  Grav.} {\bfseries 32} no.~2, (2015) 024001}.

\bibitem{countryman2019lowlatency}
S.~Countryman, A.~Keivani, I.~Bartos, Z.~Marka, T.~Kintscher, R.~Corley,
  E.~Blaufuss, C.~Finley, and S.~Marka, ``Low-latency algorithm for
  multi-messenger astrophysics (llama) with gravitational-wave and high-energy
  neutrino candidates,'' 2019.
\newblock \url{https://arxiv.org/abs/1901.05486}.

\bibitem{Keivani:2019Mx}
A.~Keivani, D.~Veske, S.~Countryman, I.~Bartos, K.~R. Corely, Z.~Marka, and
  S.~Marka, \href{http://dx.doi.org/10.22323/1.358.0930}{``{Multi-messenger
  Gravitational-Wave + High-Energy Neutrino Searches with LIGO, Virgo and
  IceCube},''} in {\em Proceedings of 36th International Cosmic Ray Conference
  {\textemdash} PoS(ICRC2019)}, vol.~358, p.~930.
\newblock 2019.
\newblock \href{http://arxiv.org/abs/1908.04996}{{\ttfamily arXiv:1908.04996
  [astro-ph.HE]}}.

\bibitem{Hussain:2019gQ}
R.~Hussain, J.~Vandenbroucke, and J.~Wood,
  \href{http://dx.doi.org/10.22323/1.358.0918}{``{A Search for IceCube
  Neutrinos from the First 33 Detected Gravitational Wave Events},''} in {\em
  Proceedings of 36th International Cosmic Ray Conference {\textemdash}
  PoS(ICRC2019)}, vol.~358, p.~918.
\newblock 2019.
\newblock \href{http://arxiv.org/abs/1908.07706}{{\ttfamily arXiv:1908.07706
  [astro-ph.HE]}}.

\bibitem{PhysRevD.100.083017}
I.~Bartos, D.~Veske, A.~Keivani, Z.~M\'arka, S.~Countryman, E.~Blaufuss,
  C.~Finley, and S.~M\'arka
  \href{http://dx.doi.org/10.1103/PhysRevD.100.083017}{{\em Phys. Rev. D}
  {\bfseries 100} (Oct, 2019) 083017}.

\bibitem{Kintscher:2016uqh}
{\bfseries IceCube} Collaboration, T.~Kintscher
  \href{http://dx.doi.org/10.1088/1742-6596/718/6/062029}{{\em J. Phys. Conf.
  Ser.} {\bfseries 718} no.~6, (2016) 062029}.

\bibitem{BARET20111}
B.~Baret, I.~Bartos, B.~Bouhou, A.~Corsi, I.~D. Palma, C.~Donzaud, V.~V.
  Elewyck, C.~Finley, G.~Jones, A.~Kouchner, S.~Márka, Z.~Márka, L.~Moscoso,
  E.~Chassande-Mottin, M.~A. Papa, T.~Pradier, P.~Raffai, J.~Rollins, and
  P.~Sutton
  \href{http://dx.doi.org/https://doi.org/10.1016/j.astropartphys.2011.04.001}{{\em
  Astroparticle Physics} {\bfseries 35} no.~1, (2011) 1 -- 7}.

\bibitem{Fang:2017tla}
K.~Fang and B.~D. Metzger
  \href{http://dx.doi.org/10.3847/1538-4357/aa8b6a}{{\em Astrophys. J.}
  {\bfseries 849} no.~2, (2017) 153}.

\bibitem{Decoene:2019eux}
V.~Decoene, C.~Gu\'epin, K.~Fang, K.~Kotera, and B.~D. Metzger
  \href{http://dx.doi.org/10.1088/1475-7516/2020/04/045}{{\em JCAP} {\bfseries
  04} (2020) 045}.

\bibitem{LIGOScientific:2018mvr}
{\bfseries LIGO Scientific, Virgo} Collaboration, B.~P. Abbott {\em et~al.}
  \href{http://dx.doi.org/10.1103/PhysRevX.9.031040}{{\em Phys. Rev. X}
  {\bfseries 9} no.~3, (2019) 031040}.

\bibitem{Keivani_2021}
A.~Keivani, J.~A. Kennea, P.~A. Evans, A.~Tohuvavohu, R.~Rapisura, S.~R. Oates,
  S.~Countryman, I.~Bartos, Z.~M{\'{a}}rka, D.~Veske, S.~M{\'{a}}rka, and D.~B.
  Fox \href{http://dx.doi.org/10.3847/1538-4357/abdab4}{{\em The Astrophysical
  Journal} {\bfseries 909} no.~2, (Mar, 2021) 126}.

\bibitem{Abbott:2020khf}
{\bfseries LIGO Scientific, Virgo} Collaboration, R.~Abbott {\em et~al.}
  \href{http://dx.doi.org/10.3847/2041-8213/ab960f}{{\em Astrophys. J. Lett.}
  {\bfseries 896} no.~2, (2020) L44}.

\end{thebibliography}\endgroup

\clearpage
\section*{Full Author List: IceCube Collaboration}

\scriptsize
\noindent
R. Abbasi$^{17}$,
M. Ackermann$^{59}$,
J. Adams$^{18}$,
J. A. Aguilar$^{12}$,
M. Ahlers$^{22}$,
M. Ahrens$^{50}$,
C. Alispach$^{28}$,
A. A. Alves Jr.$^{31}$,
N. M. Amin$^{42}$,
R. An$^{14}$
% ...
\\

\noindent
% ...
$^{12}$ Universit{\'e} Libre de Bruxelles, Science Faculty CP230, B-1050 Brussels, Belgium \\
% ...
$^{14}$ Department of Physics and Laboratory for Particle Physics and Cosmology, Harvard University, Cambridge, MA 02138, USA \\
% ...
$^{17}$ Department of Physics, Loyola University Chicago, Chicago, IL 60660, USA \\
$^{18}$ Dept. of Physics and Astronomy, University of Canterbury, Private Bag 4800, Christchurch, New Zealand \\
% ...
$^{22}$ Niels Bohr Institute, University of Copenhagen, DK-2100 Copenhagen, Denmark \\
% ...
$^{28}$ D{\'e}partement de physique nucl{\'e}aire et corpusculaire, Universit{\'e} de Gen{\`e}ve, CH-1211 Gen{\`e}ve, Switzerland \\
% ...
$^{31}$ Karlsruhe Institute of Technology, Institute for Astroparticle Physics, D-76021 Karlsruhe, Germany  \\
% ...
$^{42}$ Bartol Research Institute and Dept. of Physics and Astronomy, University of Delaware, Newark, DE 19716, USA \\
% ...
$^{50}$ Oskar Klein Centre and Dept. of Physics, Stockholm University, SE-10691 Stockholm, Sweden \\
% ...
$^{59}$ DESY, D-15738 Zeuthen, Germany \\
% ...

\subsection*{Acknowledgements}

\noindent
USA {\textendash} U.S. National Science Foundation-Office of Polar Programs,
U.S. National Science Foundation-Physics Division,
U.S. National Science Foundation-EPSCoR,
Wisconsin Alumni Research Foundation,
Center for High Throughput Computing (CHTC) at the University of Wisconsin{\textendash}Madison,
Open Science Grid (OSG),
Extreme Science and Engineering Discovery Environment (XSEDE),
Frontera computing project at the Texas Advanced Computing Center,
U.S. Department of Energy-National Energy Research Scientific Computing Center,
Particle astrophysics research computing center at the University of Maryland,
Institute for Cyber-Enabled Research at Michigan State University,
and Astroparticle physics computational facility at Marquette University;
Belgium {\textendash} Funds for Scientific Research (FRS-FNRS and FWO),
FWO Odysseus and Big Science programmes,
and Belgian Federal Science Policy Office (Belspo);
Germany {\textendash} Bundesministerium f{\"u}r Bildung und Forschung (BMBF),
Deutsche Forschungsgemeinschaft (DFG),
Helmholtz Alliance for Astroparticle Physics (HAP),
Initiative and Networking Fund of the Helmholtz Association,
Deutsches Elektronen Synchrotron (DESY),
and High Performance Computing cluster of the RWTH Aachen;
Sweden {\textendash} Swedish Research Council,
Swedish Polar Research Secretariat,
Swedish National Infrastructure for Computing (SNIC),
and Knut and Alice Wallenberg Foundation;
Australia {\textendash} Australian Research Council;
Canada {\textendash} Natural Sciences and Engineering Research Council of Canada,
Calcul Qu{\'e}bec, Compute Ontario, Canada Foundation for Innovation, WestGrid, and Compute Canada;
Denmark {\textendash} Villum Fonden and Carlsberg Foundation;
New Zealand {\textendash} Marsden Fund;
Japan {\textendash} Japan Society for Promotion of Science (JSPS)
and Institute for Global Prominent Research (IGPR) of Chiba University;
Korea {\textendash} National Research Foundation of Korea (NRF);
Switzerland {\textendash} Swiss National Science Foundation (SNSF);
United Kingdom {\textendash} Department of Physics, University of Oxford.

\end{document}